\documentclass[pra,aps,twocolumn,showpacs]{revtex4} 
\newcommand{\ket}[1]{|#1\rangle} 
\newcommand{\bra}[1]{\langle #1|} 
\newcommand{\Tr}{\text{Tr}} 
\begin{document}
\title{Quantal interferometry with dissipative internal motion}
\author{Erik Sj\"{o}qvist\footnote{Electronic address: 
eriks@kvac.uu.se}} 
\affiliation{Department of Quantum Chemistry, 
Uppsala University, Box 518, Se-751 20 Uppsala, Sweden}
\date{\today}       
\begin{abstract}
In presence of dissipation, quantal states may acquire complex-valued
phase effects. We suggest a notion of dissipative interferometry that
accommodates this complex-valued structure and that may serve as a
tool for analyzing the effect of certain kinds of external influences
on quantal interference. The concept of mixed-state phase and
concomitant gauge invariance is extended to dissipative internal
motion. The resulting complex-valued mixed-state interference effects
lead to well-known results in the unitary limit and in the case of
dissipative motion of pure quantal states. Dissipative interferometry 
is applied to fault-tolerant geometric quantum computation.
\end{abstract}
\pacs{03.65.Vf, 03.67.Lx, 03.75.Dg}
\maketitle
\section{Introduction}
Garrison and Wright \cite{garrison88} were first to address the effect 
of dissipation on quantal interference, with particular emphasis on the 
geometric phase in cyclic motion. By modeling the dissipative quantal 
motion with non-Hermitian Hamiltonians, they arrived at a complex-valued 
phase concept, the geometric interpretation of which was formulated 
in a biorthonormal description \cite{lancaster85}. This result has 
triggered further work on complex-valued geometric phase effects
\cite{chu89,aitchison92,gao92,massar96,mostafazadeh99,keck03}. 

In this paper, we revisit, from the point of view of interferometry,
the concept of phase in the presence of dissipation. Our point of
departure is the concept of gauge invariance applied to interference
in dissipative systems. We focus on a proper treatment of gauge
invariance of the non-Hermitian description of dissipative
interferometry. When the interfering system carries some internal
degrees of freedom, consideration of this gauge symmetry is shown to
lead to a complex-valued geometric phase effect for arbitrary input
states. This effect constitutes the non-Hermitian generalization 
of the mixed-state geometric phase put forward in Ref. \cite{sjoqvist00} 
and, in the particular case of pure cyclic internal states, it 
reduces to the Garrison-Wright phase \cite{garrison88}.

Geometric quantum computation, first proposed in Ref. \cite{zanardi99}
and experimentally demonstrated in Ref. \cite{jones00}, has attracted
considerable interest recently due to its predicted resilience to
certain kinds of errors. This attractive feature has been analyzed
from different perspectives in the Abelian case, such as random
unitary perturbations \cite{ekert00,zhu04} and decoherence \cite{carollo04}
(for similar analyses of non-Abelian geometric quantum computation,
see Refs. \cite{recati02,pachos02,cen04,fuentes03,solinas03}). We
demonstrate fault-tolerance with respect to dissipative decay for an
Abelian geometric phase shift gate.

The biorthonormal approach to dissiptative quantal motion is described
in the next section. In particular, we put forward an extension to the
mixed-state case, in order to pave the way for the subsequent analysis
in Sec. III of dissipative interferometry with internal degrees of
freedom. Section IV contains an analysis of a nonadiabatic one-qubit
geometric phase shift gate undergoing dissipative decay. The paper
ends with the conclusions.

\section{Biorthonormal approach}
In presence of dissipation, we expect the norm of state vectors to
change in time. To illustrate this, consider a dissipative system
modeled by a partial absorber, characterized by the transmission
probability $0<T\leq 1$. When passing through the absorber, any
$\ket{\psi}$ in Hilbert space $\mathcal{H}$ transforms as $\ket{\psi}
\rightarrow \sqrt{T}\ket{\psi}$, which has norm reduced by a factor 
$T$. On the other hand, the space of pure states is $P(\mathcal{H}) =
\mathcal{H}/({\bf C}-\{0\})$ \cite{aitchison92,pati95}, 
${\bf C}-\{ 0 \}$ being the set of nonzero complex numbers, i.e.,
$\ket{\psi}$ and $\sqrt{T} \ket{\psi}$ should be regarded as the 
same state.  

To deal with the projective structure $P(\mathcal{H})$, one defines
pure states as generalized one dimensional projectors \cite{chu89}
that take the form
\begin{eqnarray} 
P & = & \ket{\alpha} \bra{\beta} , 
\end{eqnarray} 
such that $\Tr P = \bra{\beta} \alpha \rangle =1$. Here $\ket{\alpha}, 
\ket{\beta} \in \mathcal{H}$ are said to be binormalized. 
Dissipation may be modeled by a time-dependent non-Hermitian
Hamiltonian $H(t)$, which we assume to have a nondegenerate
complex-valued discrete spectrum. In terms of the linear operators 
$L(t)$ and $R(t)$, being solutions of ($\hbar =1$ from now on)
\begin{eqnarray} 
i\dot{L}(t) & = & H(t) L(t) ,   
\nonumber \\ 
i\dot{R}(t) & = & H^{\dagger} (t) R(t)  
\label{eq:eqm} 
\end{eqnarray}
with $L (0) = R (0) =I$, pure states evolve as 
\begin{eqnarray} 
P \rightarrow P(t) = L(t) P R^{\dagger} (t) ,  
\end{eqnarray}
which is trace preserving for arbitrary $P$ provided that 
\begin{eqnarray} 
R^{\dagger} (t) L(t) = I . 
\label{eq:norm}  
\end{eqnarray}
Indeed, we obtain from Eq. (\ref{eq:eqm}) that 
\begin{eqnarray} 
\frac{d}{dt} \Big( R^{\dagger} (t) L(t) \Big) = 0, 
\end{eqnarray}
which together with $L (0) = R (0) =I$ implies Eq. (\ref{eq:norm}). 
In the example with a partial absorber discussed above, we may put  
$L = \sqrt{T}$ and $R = 1/\sqrt{T}$ so that $\ket{\psi}\bra{\psi} 
\rightarrow L\ket{\psi}\bra{\psi}R^{\dagger} = \ket{\psi}\bra{\psi}$, 
in concurrence with the projective structure $P(\mathcal{H})$. 

In many physical situations pure states do not provide an accurate
state description and one has to resort to mixed states. To deal  
with these cases, we introduce the generalized density operator
\begin{eqnarray} 
\rho = \sum_{k=1}^N w_k \ket{\alpha_k} \bra{\beta_k} ,   
\end{eqnarray} 
where $w_k\geq 0$ are real-valued and sum up to unity so that $\Tr
\rho =1$ by requiring $\bra{\beta_k}\alpha_l\rangle=\delta_{kl}$ 
\cite{remark1}.  The set $\{\ket{\alpha_k},\ket{\beta_k}; 
k=1,\ldots N \}$ with $\sum_{k=1}^N \ket{\alpha_k} \bra{\beta_k} = 
I$ is said to be a biorthonormal complete basis of the $N$ 
dimensional Hilbert space $\mathcal{H}$, such that $\ket{\alpha_k}$ 
and $\ket{\beta_k}$ are eigenvectors of $\rho$ and $\rho^{\dagger}$, 
respectively. $\rho$ evolves as
\begin{eqnarray} 
\rho \rightarrow \rho(t) = L(t) \rho R^{\dagger} (t) ,  
\end{eqnarray}
which in conjunction with Eq. (\ref{eq:norm}) assures preserved
trace. It further follows that $\{ \ket{\alpha_k (t)}
\equiv L(t) \ket{\alpha_k} \}$ and $\{ \ket{\beta_k (t)} 
\equiv R(t) \ket{\beta_k} \}$ are nonorthogonal sets of 
eigenvectors of $\rho (t)$ and $\rho^{\dagger} (t)$, respectively, 
both with time-independent semi-positive eigenvalues $w_k$. 

\section{Dissipative interferometry}
\begin{figure}[ht]
\begin{center}
\setlength{\unitlength}{1mm}
\begin{picture}(80,50)(0,25)
\put(15,30){\line(0,1){25}} 
\put(15,30){\line(1,0){40}} 
\put(15,55){\line(1,0){45}}
\put(55,30){\line(0,1){25}}
\put(55,55){\vector(1,0){7}}
\put(55,55){\vector(0,1){7}} 
\put(55,55){\line(1,0){10}}
\put(55,55){\line(0,1){10}} 
\put(5,30){\vector(1,0){6}}
\put(5,30){\line(1,0){10}}
\put(-6,38){\makebox(30,6){$\ket{1}$}}
\put(15,22){\makebox(30,6){$\ket{0}$}}
\put(40,66){\makebox(30,6){$\ket{1}$}}
\put(55,52){\makebox(30,6){$\ket{0}$}}
\put(15,57){\makebox(30,6){{$\ket{0}$}}} 
\put(45,38){\makebox(30,6){{$\ket{1}$}}}
\put(-15,27){\makebox(30,6){{$\ket{0}$}}}
\put(13,28){\line(1,1){4}} \put(53,53){\line(1,1){4}}
{\thicklines \put(13,53){\line(1,1){4}} \put(53,28){\line(1,1){4}}}
\put(15,45){\circle*{2}} 
\put(8,42){\makebox(30,6){{$\sqrt{T}e^{i\chi}$}}} 
\end{picture}
\end{center}
\caption{Interferometer setup illustrating the effect of partial 
absorption and phase shift. $1-T$ is the absorption probability 
and $\chi$ a variable U(1) phase.}
\label{fig:fig1}
\end{figure}
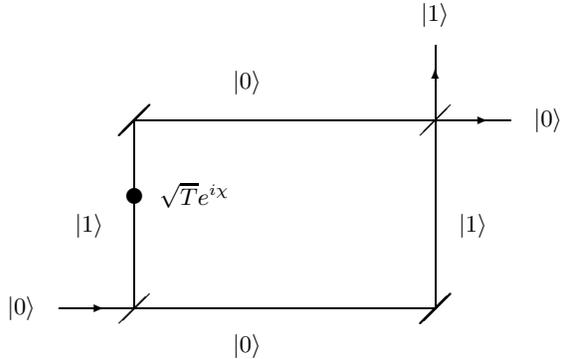

Consider a single beam of particles incident on the standard
Mach-Zehnder interferometer shown in Fig. \ref{fig:fig1}. At each
equal-time slice in the interferometer, $\ket{0},\ket{1}$ span the
Hilbert space $\mathcal{H}_s$ describing the spatial beam-pair.
All horizontal (vertical) beams are denoted $\ket{0}$
($\ket{1}$). Let $1-T$, $0<T\leq 1$, be the absorption probability of
a static partial absorber in the $\ket{1}$ beam added to it a variable
U(1) phase $\chi$. What is the output intensity consistent with the
biorthonormal description?

To appreciate what follows, we may first recall that the standard 
analysis \cite{rauch84} yields the output intensity
\begin{eqnarray} 
\widetilde{\mathcal{I}}_{[0]} \propto 
1+\frac{2\sqrt{T}}{1+T} \cos \chi 
\label{eq:standard}
\end{eqnarray}
in the $\ket{0}$ output channel. It is perhaps tempting to interpret
this result in terms of an interference pattern characterized by
visibility $\widetilde{\nu} = 2\sqrt{T}/(1+T)$ and phase shift
$\chi$. On the other hand, as noted in the previous section, the state
along the $\ket{1}$ beam inside the interferometer is unaffected by
the application of the U(1) phase, as well as of the partial
absorber. This gauge symmetry suggests that a mere multiplication of
$\sqrt{T} e^{i\chi}$ in the $\ket{1}$ beam should only change the
phase in the output, while the visibility should remain unity. We now
show that the biorthonormal formalism accommodates a natural notion of
phase and visibility, adapted to this intuition.

Let $z=\sqrt{T} e^{i\chi}$ and identify $L = z$ and $R = 1/z^{\ast}$ 
so that $\ket{\psi}\bra{\psi} \rightarrow L \ket{\psi} \bra{\psi} 
R^{\dagger} = \ket{\psi} \bra{\psi}$ for any $\ket{\psi} \in 
\mathcal{H}_s$. In other words, $z$ does not affect the local 
motion in the $\ket{1}$ beam, but it may show up in interference. 
To verify this latter point, we note that the effect of $z$ may 
be represented by the operators 
\begin{eqnarray}
\widetilde{L} & = & \ket{0} \bra{0} + z \ket{1} \bra{1} , 
\nonumber \\ 
\widetilde{R} & = & \ket{0} \bra{0} + \frac{1}{z^{\ast}} \ket{1} \bra{1} . 
\end{eqnarray}
By analyzing the interferometer, we obtain that the input state 
$\ket{0} \bra{0}$ transforms as  
\begin{eqnarray} 
\ket{0} \bra{0} & \rightarrow & \textrm{H} \sigma_x \widetilde{L} \textrm{H} 
\ket{0} \bra{0} \textrm{H}^{\dagger} 
\widetilde{R}^{\dagger} \sigma_x \textrm{H}^{\dagger} 
\nonumber \\ 
 & = & \frac{1}{4} (2 + \frac{1}{z} + z) \ket{0} \bra{0}   
\nonumber \\ 
 & & + \frac{1}{4} (2 - \frac{1}{z} - z) \ket{1} \bra{1}  
\nonumber \\ 
 & & + \textrm{interference terms} 
\end{eqnarray}
with $\sigma_x$ and $\textrm{H}$ the standard Pauli-X and Hadamard 
operator, respectively, acting on $\mathcal{H}_s$. In the $\ket{0}$ 
output channel, we obtain the $z$-dependent complex-valued intensity 
\begin{eqnarray} 
\mathcal{I}_{[0]} \propto 2 + \frac{1}{z} + z .  
\end{eqnarray}
This intensity displays a singularity at the origin in the complex 
$z$ plane. Physically, this singular point at $T=0$ corresponds to 
vanishing interference or, equivalently, a situation where the path 
of the particles is perfectly known. The complex-valued phase shift 
$\phi$ and visibility $\nu$ are defined as 
\begin{eqnarray} 
\mathcal{I}_{[0]} \propto 1 + \nu \cos \phi , 
\label{eq:complexintensity}
\end{eqnarray}
which yields 
\begin{eqnarray}
e^{i\phi} & = & z , 
\nonumber \\ 
\nu & = & 1,  
\end{eqnarray} 
as desired. 

To further analyze the relation between $\mathcal{I}_{[0]}$ and the 
experimental parameters $\chi$ and $T$, let us focus on the 
complex-valued interference term $\cos \phi \equiv \mathcal{J}_{[0]}$. 
By introducing the polar decomposition $\mathcal{J}_{[0]} =  
|\mathcal{J}_{[0]}| e^{-i\vartheta}$, we obtain 
\begin{eqnarray} 
\tan \vartheta & = & \frac{1-T}{1+T} \tan \chi , 
\nonumber \\ 
|\mathcal{J}_{[0]}| & = & \sqrt{\cos^2 \chi + \frac{(1-T)^2}{4T}} ,  
\end{eqnarray}
which shows that $\mathcal{J}_{[0]}$ rotates in the complex plane 
with angular frequency that tends to $\chi$ in the singular 
$T\rightarrow 0$ limit where $|\mathcal{J}_{[0]}|$ becomes infinite. 
In the unitary limit $T\rightarrow 1$, it follows directly from 
Eq. (\ref{eq:complexintensity}) that $\mathcal{J}_{[0]}$ oscillates 
along the real axis according to the expected $\cos \chi$.  

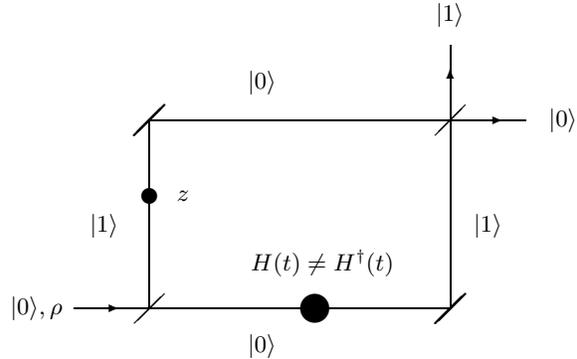
\begin{figure}[ht]
\begin{center}
\setlength{\unitlength}{1mm}
\begin{picture}(80,50)(0,25)
\put(15,30){\line(0,1){25}} 
\put(15,30){\line(1,0){40}} 
\put(15,55){\line(1,0){45}}
\put(55,30){\line(0,1){25}}
\put(55,55){\vector(1,0){7}}
\put(55,55){\vector(0,1){7}} 
\put(55,55){\line(1,0){10}}
\put(55,55){\line(0,1){10}} 
\put(5,30){\vector(1,0){6}}
\put(5,30){\line(1,0){10}}
\put(-6,38){\makebox(30,6){$\ket{1}$}}
\put(15,22){\makebox(30,6){$\ket{0}$}}
\put(40,66){\makebox(30,6){$\ket{1}$}}
\put(55,52){\makebox(30,6){$\ket{0}$}}
\put(15,57){\makebox(30,6){{$\ket{0}$}}} 
\put(45,38){\makebox(30,6){{$\ket{1}$}}}
\put(-15,27){\makebox(30,6){{$\ket{0},\rho$}}}
\put(13,28){\line(1,1){4}} \put(53,53){\line(1,1){4}}
{\thicklines \put(13,53){\line(1,1){4}} \put(53,28){\line(1,1){4}}}
\put(15,45){\circle*{2}} 
\put(37,30){\circle*{4}} 
\put(4.5,42){\makebox(30,6){{$z$}}} 
\put(23,33){\makebox(30,6){{$H(t) \neq H^{\dagger}(t)$}}} 
\end{picture}
\end{center}
\caption{Interferometer setup illustrating the interference effect 
due to dissipative internal motion.  The dissipative motion is modeled
by the non-Hermitian Hamiltonian $H(t)$ that acts on the internal
degrees of freedom of the interfering particles. $z$ is a nonzero
complex number.}
\label{fig:fig2}
\end{figure}

Next, we extend the above setup and assume that the particles carry 
some internal degrees of freedom prepared in an input state that is 
described by the generalized density operator $\rho$, so that the 
incoming state is characterized by the generalized density operator 
$\varrho_{{\textrm{in}}} = \ket{0} \bra{0} \otimes \rho$ acting on 
the full Hilbert space $\mathcal{H}_s \otimes \mathcal{H}$. Suppose 
that
\begin{eqnarray}
{\bf L} & = & \ket{0} \bra{0} \otimes L + 
z \ket{1} \bra{1} \otimes I , 
\nonumber \\ 
{\bf R} & = & \ket{0} \bra{0} \otimes R + 
\frac{1}{z^{\ast}} \ket{1} \bra{1} \otimes I  
\end{eqnarray}
with $L,R$ being solutions of Eq. (\ref{eq:eqm}) for 
$H(t)\neq H^{\dagger}(t)$, are applied between the first beam-splitter 
and the mirror pair, see Fig. \ref{fig:fig2}, and the complex-valued 
$z$ must in general contain absorption in order to fully exploit the 
complex-valued structure of the phase shift resulting from the 
non-Hermitian transformation of the internal motion. The output 
state becomes
\begin{eqnarray} 
\varrho_{{\textrm{out}}} = {\bf U}_{B} {\bf U}_{M} {\bf L} {\bf U}_{B} 
\varrho_{{\textrm{in}}} {\bf U}_{B}^{\dagger} 
{\bf R}^{\dagger} {\bf U}_{M}^{\dagger} 
{\bf U}_{B}^{\dagger}  
\label{eq:output}
\end{eqnarray}  
with ${\bf U}_{M} = \sigma_x \otimes I$ and ${\bf U}_{B} = {\textrm{H}}
\otimes I$, where $\sigma_x$ and ${\textrm{H}}$ as above acting on 
$\mathcal{H}_s$. The operators ${\bf U}_{M}$, ${\bf U}_{B}$, ${\bf L}$, 
and ${\bf R}$ act on the full Hilbert space $\mathcal{H}_s \otimes
\mathcal{H}$. ${\bf L}$ and ${\bf R}$ correspond to the application 
of $L$ and $R$ along the $\ket{0}$ path and the $z$ operation similarly 
along $\ket{1}$. Direct evaluation yields the output state  
\begin{eqnarray} 
\varrho_{{\textrm{out}}} & = & \frac{1}{4} \ket{0} \bra{0} \otimes 
\Big( L \rho R^{\dagger} + \rho + 
\frac{1}{z} L \rho + z \rho R^{\dagger} \Big) 
\nonumber \\ 
 & & + \frac{1}{4} \ket{1} \bra{1} \otimes \Big( L \rho R^{\dagger} + 
\rho - \frac{1}{z} L \rho - z \rho R^{\dagger} \Big) 
\nonumber \\ 
 & & + {\textrm{interference terms}} . 
\end{eqnarray}
In the $\ket{0}$ output channel, we thus obtain 
\begin{eqnarray} 
\mathcal{I}_{[0]} \propto 2+\frac{1}{z} \Tr \big[ L \rho \big] + 
z \Tr \big[ \rho R^{\dagger} \big] . 
\label{eq:0intensity1}
\end{eqnarray}

In analogy with Eq. (\ref{eq:complexintensity}), we may now introduce the
complex-valued relative phase $\Phi$ and visibility $\mathcal{V}$ as
\begin{eqnarray} 
\mathcal{I}_{[0]} \propto 1+\mathcal{V} \cos \big[ \phi - \Phi \big] ,   
\label{eq:0intensity2}
\end{eqnarray}
with $z=e^{i\phi}\neq 0$, $\phi \in {\bf C}$. Comparing Eqs. 
(\ref{eq:0intensity1}) and (\ref{eq:0intensity2}), yields 
\begin{eqnarray} 
e^{i\Phi} & = & 
\sqrt{\frac{\Tr \big[ L \rho \big]}
{\Tr \big[ \rho R^{\dagger} \big]}} , 
\nonumber \\ 
\mathcal{V} & = & \sqrt{\Tr \big[ L \rho \big] 
\Tr \big[ \rho R^{\dagger} \big]} ,  
\label{eq:pancharatnam}
\end{eqnarray}
which is the desired complex-valued generalization of relative phase 
and interference visibility put forward in Ref. \cite{sjoqvist00}.  
In the case where $H(t)$ is Hermitian, we have that $L = R \equiv U$ 
is unitary and we obtain 
\begin{eqnarray} 
e^{i\Phi} & = & \sqrt{\frac{\Tr \big[ U \rho \big]}
{\Tr \big[ \rho U^{\dagger} \big]}} , 
\nonumber \\ 
\mathcal{V} & = & \sqrt{\Tr \big[ U \rho \big] 
\Tr \big[ \rho U^{\dagger} \big]} ,  
\end{eqnarray}
which is consistent with Ref. \cite{sjoqvist00} in the case where 
$\rho$ is Hermitian. Note, however, that $\Phi$ may be complex-valued 
in the general unitary case where $\ket{\beta_k} \neq \ket{\alpha_k}$. 

We now introduce a concept of geometric phase associated with 
the path $\mathcal{C}:t\in[0,\tau] \rightarrow \rho(t) = L (t) 
\rho R^{\dagger} (t)$. The basic observation for this purpose 
is that there is an equivalence set $\mathcal{A}$ of operator pairs 
$\widetilde{L} (t),\widetilde{R} (t)$ that all generate 
$\mathcal{C}$, namely those of the form 
\begin{eqnarray} 
\widetilde{L} (t) & = & 
L (t) \sum_{k=1}^N z_k (t) \ket{\alpha_k} \bra{\beta_k} , 
\nonumber \\ 
\widetilde{R} (t) & = & 
R (t) \sum_{k=1}^N \frac{1}{z_k^{\ast}(t)} \ket{\beta_k} \bra{\alpha_k} , 
\label{eq:equivset}
\end{eqnarray}
where $z_k(t),t \in [0,\tau]$, are nonvanishing complex numbers such
that $z_k(0)=1$. The equivalence set $\mathcal{A}$ is the proper
non-Hermitian generalization of that in the unitary case introduced in
Ref. \cite{singh03}. The existence of $\mathcal{A}$ expresses the
gauge symmetry of dissipative motion of mixed quantal states. We may
identify $\{ L^{\parallel} (t), R^{\parallel} (t) \} \in \mathcal{A}$
fulfilling
\begin{eqnarray} 
\bra{\beta_k} R^{\parallel \dagger} (t) \dot{L}^{\parallel} (t) 
\ket{\alpha_k} = 0 , \ k=1,\ldots,N. 
\label{eq:parallel}
\end{eqnarray} 
These constitute parallel transport conditions in a fiber 
bundle with structure group $({\bf C}-\{ 0 \})^N$. Substituting 
$L^{\parallel} (t),R^{\parallel} (t) = \widetilde{L} (t),
\widetilde{R} (t)$, with $\widetilde{L} (t),\widetilde{R} (t)$ 
given by Eq. (\ref{eq:equivset}), into Eq. (\ref{eq:parallel}), 
we obtain
\begin{eqnarray} 
z_k^{\parallel}  (\tau) = 
e^{-\int_0^{\tau} \bra{\beta_k} R^{\dagger} (t) 
\dot{L} (t) \ket{\alpha_k} dt} = 
e^{-\int_0^{\tau} \bra{\beta_k (t)} \dot{\alpha}_k (t) \rangle dt},
\end{eqnarray} 
where we have used that $z_k^{\parallel} (0)=1$. Putting this into 
the relative phase, we identify the mixed-state generalization of 
the complex-valued geometric phase factor as 
\begin{eqnarray} 
 & & e^{i\Gamma} =  
\sqrt{\frac{\Tr \big[ L^{\parallel} (\tau) \rho \big]}
{\Tr \big[ \rho R^{\parallel \dagger} (\tau) \big]}} 
\nonumber \\ 
 & = & \sqrt{\frac{\sum_{k=1}^N w_k \bra{\beta_k} L(\tau) 
\ket{\alpha_k} e^{-\int_0^{\tau} \bra{\beta_k (t)}  
\dot{\alpha}_k (t) \rangle dt}}{\sum_{k=1}^N w_k \bra{\beta_k} 
R^{\dagger}(\tau) \ket{\alpha_k} e^{\int_0^{\tau} \bra{\beta_k (t)} 
\dot{\alpha}_k (t) \rangle dt}}} . 
\nonumber \\ 
\label{eq:gp} 
\end{eqnarray}
We may verify that $e^{i\Gamma}|_{\widetilde{L} (t),\widetilde{R} (t)} = 
e^{i\Gamma}|_{L(t),R(t)}$ for any $\widetilde{L} (t), \widetilde{R} (t) 
\in \mathcal{A}$. Thus, $e^{i\Gamma}$ is a property of $\mathcal{C}$.  

Let us now analyse two important special cases. First, assume 
that $H(t)$ is Hermitian, so that $L(t)=R(t)\equiv U(t)$ is a 
one-parameter family of unitarities. In this case, Eq. (\ref{eq:gp}) 
takes the form 
\begin{eqnarray} 
e^{i\Gamma} & = & \sqrt{\frac{\sum_{k=1}^N w_k \bra{\beta_k} U(\tau) 
\ket{\alpha_k}  e^{-\int_0^{\tau} \bra{\beta_k (t)}  
\dot{\alpha}_k (t) \rangle dt}}{\sum_{k=1}^N w_k \bra{\beta_k} 
U^{\dagger}(\tau) \ket{\alpha_k} e^{\int_0^{\tau} \bra{\beta_k (t)}  
\dot{\alpha}_k (t) \rangle dt}}} , 
\nonumber \\  
\label{eq:unitary}
\end{eqnarray}
where $\ket{\alpha_k (t)} = U(t) \ket{\alpha_k}$ and 
$\ket{\beta_k (t)} = U(t) \ket{\beta_k}$. Equation (\ref{eq:unitary}) 
is consistent with the mixed-state geometric phase in unitary 
evolution \cite{sjoqvist00} when $\rho$ is Hermitian, but $\Gamma$ 
may still be complex-valued in the general unitary case where 
$\ket{\beta_k} \neq \ket{\alpha_k}$. 

Secondly, in the pure cyclic case, defined by $\rho = \ket{\alpha}
\bra{\beta}$, $L(\tau) \ket{\alpha} = e^{i\zeta} \ket{\alpha}$, and $R
(\tau) \ket{\beta} = e^{i\zeta^{\ast}} \ket{\beta}$ with $\zeta$ some
complex number \cite{garrison88}, we obtain
\begin{eqnarray} 
e^{i\Gamma} & = & e^{i\zeta} e^{-\int_0^{\tau} \bra{\beta (t)} 
\dot{\alpha} (t) \rangle dt} .  
\end{eqnarray}
In analogy with Ref. \cite{aharonov87}, we may put 
\begin{eqnarray}
\ket{\widetilde{\alpha} (t)} & = & e^{-if(t)} \ket{\alpha(t)},  
\nonumber \\ 
\ket{\widetilde{\beta} (t)} & = & e^{-if^{\ast}(t)} \ket{\beta(t)}
\end{eqnarray} 
such that $f(\tau)-f(0)=\zeta$. This yields 
\begin{eqnarray}
\Gamma & = & i\int_0^{\tau} \bra{\widetilde{\beta} (t)} 
\dot{\widetilde{\alpha}} (t) \rangle dt ,  
\end{eqnarray}
which is consistent with Ref. \cite{garrison88}. 

\section{Fault-tolerant geometric quantum computation}
Abelian nonadiabatic geometric quantum computation has been proposed 
\cite{xiang-bin01} in order to achieve high-speed fault-tolerant 
implementations of quantum gates. Here, we analyze the resilience of a
nonadiabatic one-qubit geometric phase shift gate to dissipation.

The physical scenario for the dissipative phase shift gate is an
unstable two-level atom interacting with an external electric field
that rotates uniformly around the $z$ axis. By using the rotating 
wave approximation, the internal quantal motion of the atom may, 
in the rotating frame, be described by the non-Hermitian Hamiltonian
\begin{eqnarray}
H=\frac{1}{2}(\eta -i\gamma) \sigma_z \equiv \frac{1}{2}\omega
\sigma_z, 
\end{eqnarray} 
where $\eta$ is the detuning, $\gamma$ is the average decay rate, and
for simplicity we have neglected small off-diagonal terms in the
weak-amplitude limit of the transverse electric field in the $x-y$
plane. Furthermore, $\sigma_z = \ket{g} \bra{g} -\ket{e}\bra{e}$,
where $\ket{g}$ and $\ket{e}$ are the unperturbed ground and first
excited atomic state, respectively. Solving Eq. (\ref{eq:eqm}) yields
\begin{eqnarray} 
L (t) & = & e^{-i\omega \sigma_z t/2} , 
\nonumber \\ 
R (t) & = & e^{-i\omega^{\ast} \sigma_z t/2} .  
\end{eqnarray}
Assume 
\begin{eqnarray} 
\rho = \frac{1+r}{2} \ket{\alpha_+} \bra{\alpha_+} + 
\frac{1-r}{2} \ket{\alpha_-} \bra{\alpha_-} , 
\end{eqnarray} 
where $0<r\leq 1$ is the nonzero length of the Bloch vector of 
the qubit and  
\begin{eqnarray} 
\ket{\alpha_+} & = & \cos \frac{\theta}{2} \ket{g} + 
\sin \frac{\theta}{2} \ket{e} , 
\nonumber \\ 
\ket{\alpha_-} & = & -\sin \frac{\theta}{2} \ket{g} + 
\cos \frac{\theta}{2} \ket{e}    
\end{eqnarray} 
make an angle $\theta \in {\bf R}$ with respect to the $z$ axis. 
In absence of decay, i.e., $\gamma = 0$, a geometric phase shift gate 
of the form 
\begin{eqnarray} 
U[\theta] = e^{i(1-\sigma_{\theta}) 2\pi (1-\cos\theta)} , 
\label{eq:gate}
\end{eqnarray} 
where $\sigma_{\theta} = \ket{\alpha_+} \bra{\alpha_+} - 
\ket{\alpha_-} \bra{\alpha_-}$, is obtained after one period 
$\tau=2\pi/\eta$, by eliminating the dynamical phase, e.g., using 
refocusing technique. The dependence upon the solid angle 
$2\pi (1-\cos\theta)$ expresses the geometric nature of $U[\theta]$.  

Now, we may notice that $\ket{\alpha_{\pm}}$ undergoes noncyclic 
evolution for all $t>0$ in the presence of decay, except in the trivial 
case where $\theta=0$. Yet, we may still compute the noncyclic 
complex-valued phases by using Eqs. (\ref{eq:pancharatnam})  and 
(\ref{eq:gp}). 

First, let $\varphi = \omega \tau$ be the complex-valued total 
precession angle. By using Eq. (\ref{eq:pancharatnam}), we may 
compute the relative phase and visibility as
\begin{eqnarray} 
\Phi & = & 
-\arctan \left[ r\cos \theta \tan \frac{\varphi}{2} \right] , 
\nonumber \\ 
\mathcal{V} & = & \sqrt{\cos^2 \frac{\varphi}{2} + 
r^2 \cos^2 \theta \sin^2 \frac{\varphi}{2}} ,  
\end{eqnarray}
which are in general complex-valued unless the average decay rate 
$\gamma$ vanishes. 

Next, by using Eq. (\ref{eq:gp}), we may compute the geometric phase 
associated with the path $t\in[0,\tau]\rightarrow L (t) \rho 
R^{\dagger} (t)$ as 
\begin{eqnarray}
\Gamma = -\arctan \left[ r \tan \frac{\Omega}{2} \right] , 
\end{eqnarray}
where 
\begin{eqnarray}
\Omega = 2\arctan \left[ \cos \theta \tan \frac{\varphi}{2} \right] - 
\varphi \cos \theta 
\end{eqnarray}
is a complex-valued analog of the geodesically closed solid angle 
on the Bloch sphere, appearing in the Hermitian case. For small 
$\gamma \tau$, we may Taylor expand $\Omega$ and obtain 
\begin{eqnarray} 
\Omega & = & 2\arctan \left[ \cos \theta \tan \frac{\eta \tau}{2} \right] 
- \eta \tau \cos \theta 
\nonumber \\ 
 & & -i\gamma \tau \cos \theta \frac{\sin^2 \frac{\eta \tau}{2} 
\sin^2 \theta}{\left( 1-\sin^2 \frac{\eta \tau}{2} 
\sin^2 \theta \right)} + \mathcal{O} \big[ (\gamma \tau)^2 \big]
\label{eq:taylor} 
\end{eqnarray}
whose real part is exactly the desired solid angle to second order in
$\gamma \tau$. This expression further entails that the one-qubit 
gate $U[\theta]$ in Eq. (\ref{eq:gate}) is fault-tolerant in $\gamma$, 
in that $\Omega$ and, since $r$ is independent of $\eta$ and $\gamma$, 
thereby also $\Gamma$ are robust to second order in the decay for 
$\tau = 2\pi/\eta$, i.e.,
\begin{eqnarray} 
\Omega & = & 2\pi (1- \cos \theta) + 
\mathcal{O} \left[ \left( \frac{2\pi\gamma}{\eta} \right)^2 \right].
\end{eqnarray}
This feature further supports the predicted resilience of geometric
quantum computation to unwanted external influences. A related result
in the context of adiabatic evolution of open quantum systems has been
found in Ref. \cite{carollo04}.

Furthermore, for small $\gamma \tau$ we may also expand the relative 
phase as 
\begin{eqnarray} 
\Phi & = & -\arctan \left[ r\cos \theta \tan \frac{\eta \tau}{2} \right] 
\nonumber \\ 
 & & - i\frac{\gamma\tau}{2} \frac{r\cos\theta}{1-(1-r^2\cos^2\theta)
\sin^2 \frac{\eta \tau}{2}} 
\nonumber \\ 
 & & + \mathcal{O} \big[ (\gamma \tau)^2 \big]. 
\end{eqnarray}
Thus, the imaginary part of $\Phi$ vanishes for $\theta = \pi/2$ but
is nonzero otherwise for all $\gamma \tau >0$. In other words, the
relative phase, which contains both dynamical and geometric
contributions, is not fault-tolerant to the dissipative decay.  This
suggests that the above demonstrated resilience to the present form of
dissipative decay is a consequence of the geometric nature of $\Gamma$,
and that a quantum gate based upon $\Phi$ would be sensitive to this
particular kind of error.

\section{Conclusions}
Two-beam interferometry with particles carrying internal degrees of
freedom has been analyzed by using a biorthonormal dynamical
description. It leads to the notion of dissipative interferometry that
may serve as a tool for analyzing the effect of certain
kinds of external influences on quantal interference. In dissipative
interferometry, phases and visibilities become complex-valued. Gauge
invariance, adapted to internal mixed input states, is defined in
terms of path invariance in state space, and has been shown to lead to
a natural concept of geometric phase that generalizes
Ref. \cite{sjoqvist00} to dissipative motion as well as
Ref. \cite{garrison88} to the mixed-state case.  Fault-tolerance of
Abelian nonadiabatic geometric quantum computation is demonstrated for
a one-qubit phase shift gate, modeled by an unstable two-level atomic
system. It would be pertinent to extend the present framework to the
non-Abelian case in order to deal with the robustness of universal
sets of quantum gates implemented by geometric means.


\begin{thebibliography}{99} 
\bibitem{garrison88} J.C. Garrison and E.M. Wright, 
Phys. Lett. A {\bf 128}, 177 (1988). 
\bibitem{lancaster85} P. Lancaster and M. Tismenetsky, 
{\it The theory of matrices} (Academic Press, San Diego, 1985), 
2nd ed.  
\bibitem{chu89} S.-I. Chu, Z.-C. Wu, and E. Layton, 
Chem. Phys. Lett. {\bf 157}, 151 (1989). 
\bibitem{aitchison92} I.J.R. Aitchison and K. Wanelik, 
Proc. R. Soc. London Ser. A {\bf 439}, 25 (1992). 
\bibitem{gao92} X.-C. Gao, J.-B. Xu, and T.-Z. Qian, 
Phys. Rev. A {\bf 46}, 3626 (1992). 
\bibitem{massar96} S. Massar, 
Phys. Rev. A {\bf 54}, 4770 (1996). 
\bibitem{mostafazadeh99} A. Mostafazadeh, 
Phys. Lett. A {\bf 264}, 11 (1999).  
\bibitem{keck03} F. Keck, H.J. Korsch, and S. Mossmann, 
J. Phys. A {\bf 36}, 2125 (2003).  
\bibitem{sjoqvist00} E. Sj\"oqvist, A.K. Pati, A. Ekert, 
J.S. Anandan, M. Ericsson, D.K.L. Oi, and V. Vedral, 
Phys. Rev. Lett. {\bf 85}, 2845 (2000). 
\bibitem{zanardi99} P. Zanardi and M. Rasetti,  
Phys. Lett. A {\bf 264}, 94 (1999). 
\bibitem{jones00} J.A. Jones, V. Vedral, A. Ekert, and 
G. Castagnoli, 
Nature (London) {\bf 403}, 869 (2000).
\bibitem{ekert00} A. Ekert, M. Ericsson, P. Hayden, H. Inamori, 
J.A. Jones, D.K.L. Oi, and V. Vedral,  
J. Mod. Opt. {\bf 47}, 2501 (2000).
\bibitem{zhu04} S.-L. Zhu and P. Zanardi, 
e-print: quant-ph/0407177. 
\bibitem{carollo04} A. Carollo, I. Fuentes-Guridi, M. Franca Santos, 
and V. Vedral, 
Phys. Rev. Lett. {\bf 92}, 020402 (2004). 
\bibitem{recati02} A. Recati, T. Calarco, P. Zanardi, J.I. Cirac, 
and P. Zoller, 
Phys. Rev. A {\bf 66}, 032309 (2002).  
\bibitem{pachos02} J. Pachos,
Phys. Rev. A {\bf 66}, 042318 (2002).
\bibitem{fuentes03} I. Fuentes-Guridi, F. Girelli, and E. Livine,
e-print: quant-ph/0311164.
\bibitem{solinas03} P. Solinas, P. Zanardi, and N. Zanghi,
Phys. Rev. A {\bf 70}, 042316 (2004).
\bibitem{cen04} L.-X. Cen and P. Zanardi, 
e-print: quant-ph/0403143. 
\bibitem{pati95} A.K. Pati, 
Phys. Lett. A {\bf 202}, 40 (1995). 
\bibitem{remark1} A weaker condition for $\Tr \rho =1$ is 
$\langle \beta_k \ket{\alpha_k} = 1$, $\forall k$. The stronger 
condition $\langle \beta_k \ket{\alpha_l} = \delta_{kl}$ plays a 
natural role in the subsequent definition of parallel transport 
in Eq. (\ref{eq:parallel}) as it is the biorthonormal analog 
to the spectral decomposition of standard (Hermitian) density 
operators.  
\bibitem{rauch84} H. Rauch and J. Summhammer, 
Phys. Lett. A {\bf 104}, 44 (1984). 
\bibitem{singh03} K. Singh, D.M. Tong, K. Basu, J.L. Chen, and J.F. Du, 
Phys. Rev. A {\bf 67}, 032106 (2003). 
\bibitem{aharonov87} Y. Aharonov and J.S. Anandan, 
Phys. Rev. Lett. {\bf 58}, 1593 (1987).
\bibitem{xiang-bin01} W. Xiang-Bin and M. Keiji, 
Phys. Rev. Lett. {\bf 87}, 097901 (2001).
\end{thebibliography}
\end{document}